\documentclass[twocolumn,preprintnumbers,amsmath,amssymb]{revtex4}
\def\be{\begin{equation}}
\def\ee{\end{equation}}
\def\bea{\begin{eqnarray}}
\def\eea{\end{eqnarray}}
\newcommand{\A}{{\mathcal{A}}}
\newcommand{\tA}{{\widetilde {\mathcal{A}}}}
\newcommand{\td}{{\widetilde d}}
\newcommand{\tal}{{\widetilde \alpha}}
\newcommand{\MSbar}{\overline{\rm MS}}
\def\bD{{\cal D}}
\def\btD{{\widetilde {\cal D}}}
\def\tK{{\widetilde K}}

\usepackage{graphicx}
\usepackage{here}

\usepackage{amssymb}

\usepackage[figuresright]{rotating}

\begin{document}


  \title{Renormalon-based resummation for QCD observables}

\author{Gorazd Cveti\v{c}}
\affiliation{Department of Physics,
Universidad T\'ecnica Federico Santa Mar\'{\i}a, Casilla 110-V, 
Valpara\'{\i}so, Chile\footnote{email: gorazd.cvetic@usm.cl; talk given by G.C.~at the 22th International Conference in Quantum Chromodynamics (QCD 19),  2-5 July 2019, Montpellier, France}}

\date{\today}

\begin{abstract}
A method of evaluation of spacelike QCD observables $\bD(Q^2)$ is presented, motivated by the renormalon structure of these quantities.
\end{abstract}

\maketitle

\section{Introduction}
\label{sec:Intr}

The theory of renormalons, and its use in the evaluation of QCD observables, has been known for some time \cite{BenekeRen}, and it has remained an area of active interest; for some new ideas that have emerged in this area more recently, see Refs.~\cite{BeJa08,MaVa,BMO,GC1,ALP}.

Borel transforms of (the leading-twist part of) the spacelike observables have specific renormalon structures, which include poles, cuts and branching points in the Borel plane. On the other hand, in the large-$\beta_0$ approximation these structures get simplified: the branching points get reduced to simple and multiple poles. In this presentation we summarize the method of Ref.~\cite{GC1} where these structures are used to evaluate spacelike QCD observables $\bD(Q^2)$. In Sec.~\ref{sec:meth} a related auxiliary quantity $\btD(Q^2)$ is introduced, which in principle contains the entire information on all the expansion coefficients of the original observable $\bD(Q^2)$, but is renomalization scale independent only at the one-loop level, and agrees with  $\bD(Q^2)$ at one-loop level. Motivated by a specific renormalization scale dependence of the Borel transform $B [\btD](b)$, a large-$\beta_0$ type of ansatz is made for $B [\btD](b)$. This leads to the correct (``dressed'') structure of the Borel transform $B [\bD](b)$ of the original observable. Subsequently, in Sec.~\ref{sec:Adl} a Neubert-type of the characteristic (distribution) function $G_D(t)$ for the original $\bD(Q^2)$ is obtained from the simple Borel transform  $B [\btD](b)$. This renormalon-based characteristic function permits evaluation (resummation) of the original observable $\bD(Q^2)$. As a specific illustration, the method is applied to the evaluation of the (leading-twist) massless Adler function and the related (timelike) decay ratio of the $\tau$ lepton semihadronic decays. At the end, the presented results are summarized.

\section{The method}
\label{sec:meth}

The perturbation expansion of the considered spacelike observable is
\begin{equation}
\bD(Q^2)_{\rm pt} = \sum_{n\geq 0} d_n(\kappa) a(\kappa Q^2)^{n+1},
\label{Dkappt}
\end{equation}
where $\mu^2 \equiv \kappa Q^2$ is the renormalization scale, and $a(\mu^2) \equiv \alpha_s(\mu^2)/\pi$. The coupling $a(\mu^2)$ satisfies the renormalization group equation (RGE)
\be
\frac{d a(\mu^2)}{d \ln \mu^2} = - \beta_0 a(\mu^2)^2 - \beta_1 a(\mu^2)^3  - \beta_2 a(\mu^2)^4 - \ldots
\label{RGE}
\ee
We can reorganize the power expansion (\ref{Dkappt}) into expansion in the logarithmic derivatives
where
\begin{equation}
{\tilde a}_{n+1}(\mu^2) \equiv \frac{(-1)^n}{\beta_0^n n!} \left( \frac{d}{d \ln \mu^2} \right)^n a(\mu^2) ,
\label{tan}
\end{equation}
 (where $n=0,1,\ldots$), which coincide with the powers $a(\mu^2)^{n+1}$ only at the one-loop level. We thus obtain the expansion
\begin{equation}
\bD(Q^2)_{\rm lpt} = \sum_{n\geq 0} {\tilde d}_n(\kappa) {\tilde a}_{n+1}(\kappa Q^2).
\label{Dkaplpt}
\end{equation}
The new expansion coefficients ${\tilde d}_n$ are unique functions of the coefficients $d_j$ ($j \leq n$), and contain all the information about them; these relations can also be inverted, and have similar structure
\begin{equation}
d_n = \sum_{s=0}^{n-1} k_s(n+1-s) \; \td_{n-s} ,
\label{dntdn}
\end{equation}
where $n=1,2,\ldots$, and $k_0(m)=0$. An auxiliary quantity $\btD$ can be introduced, which is the power expansion with the coefficients ${\tilde d}_n$
\be
\btD(Q^2;\kappa) = \sum_{n\geq 0} {\tilde d}_n(\kappa)  a(\kappa Q^2)^{n+1}.
\label{tDkap}
\ee
It has some renormalization scale ($\kappa$)-dependence when going beyond the one-loop level. The ``reorganized'' coefficients ${\tilde d}_n(\kappa)$ have a significantly simpler (one-loop-type) renormalization scale dependence than the original coefficients $d_n$
\be
\frac{d}{d \ln \kappa} \td_n(\kappa) = n \beta_0 \td_{n-1}(\kappa) \qquad
(n \geq 1),
\label{tdkap}
\ee
and $\td_0$ is $\kappa$-independent.
As a consequence, the Borel transform of the auxiliary quantity $\btD$
\be
{\rm B}[\btD](u,\kappa) = \sum_{n=0}^{\infty} \frac{\td_n(\kappa)}{n! \beta_0^n} u^n
\label{BtD1}
\ee
has the simple one-loop-type (or: large-$\beta_0$-type) renormalization scale dependence
\be
{\rm B} [\btD](u;\kappa) = \kappa^{u} {\rm B} [\btD](u) .
\label{BtDkap}
\ee
This suggests that the Borel transform ${\rm B}[\btD](u)$ has a one-loop (large-$\beta_0$) type renormalon structures (poles):
\be
{\rm B}[\btD](u) \sim 1/(p \pm u)^k, \ln(1 \pm u/p),
\label{BtDpoles}
\ee
where p and k are positive integers.
Such ans\"atze for ${\rm B}[\btD](u)$ will be used to generate the coefficients $\td_n$, and thus via Eqs.~(\ref{dntdn}) the coefficients $d_n$ of the power expansion of the full $\bD(Q^2)$ observable. However, an important question is whether these (large-$\beta_0$)-type ans\"atze for  ${\rm B}[\btD](u)$ give us correctly behaved $d_n$ coefficients of $\bD(Q^2)$, i.e., whether the Borel transform
${\rm B}[\bD](u; \kappa)$ has the (full-loop) renormalon structure expected theoretically. It can be shown numerically that this is really the case, and the reader is referred for details to Ref.~\cite{GC1}.

\section{Application to the massless Adler function}
\label{sec:Adl}
  
The Adler function $\bD(Q^2)$ is the logarithmic derivative of the quark current-current correlator. In the massless limit, the vector and axial vector channels coincide, and the perturbation expansion (\ref{Dkappt}) of this quantity is known exactly up to order $a^4$ \cite{d1,d2,d3}. Further, the leading-$\beta_0$ (LB) parts $d_n^{\rm (LB)}$ ($= \td_n^{\rm (LB)}$) of the coefficients are known to all orders $n$, and thus the LB Borel transform ${\rm B}[\bD](u)^{\rm (LB)}$ of the (massless) Adler function is known \cite{Broad}: it has simple pole ($k=1$) at $u=2$ [the leading infrared (IR) renormalon], and double poles ($k=2$) at $u=3,4,\ldots$ (IR renormalons) and at $u=-1,-2,\ldots$ [ultraviolet (UV) renormalons].

\subsection{The Borel transform of $\btD$ of Adler}
\label{subsec:BtD}

The first ansatz for the Borel ${\rm B}[\btD](u)$ includes the first two IR renormalon poles, and the first utraviolet (UV) pole $u=-1$:
\bea
\lefteqn{{\rm B}[\btD](u)^{\rm (4 P)} =}
\nonumber\\&&
\exp \left( \tK u \right) \pi {\Big \{}
\td_{2,1}^{\rm IR} \left[ \frac{1}{(2-u)} + \tal (-1) \ln \left( 1 - \frac{u}{2} \right) \right]
\nonumber\\ &&
+ \frac{ \td_{3,2}^{\rm IR} }{(3 - u)^2} + \frac{ \td_{1,2}^{\rm UV} }{(1 + u)^2} {\Big \}};
\label{BtD4P}
\eea 
which has four parameters: $\tK$, $\td_{2,1}^{\rm IR}$, $\td_{3,2}^{\rm IR}$ and $\td_{1,2}^{\rm UV}$. The values of these four parameters can be determined by requiring that the values of the first four (exactly known) perturbation expansion coefficients $d_n$ ($n=0,1,2,3$) be correctly reproduced.

In practice, this ansatz is made in a specific renormalization scheme, the Lambert MiniMOM (LMM)\footnote{LMM \cite{4l3dAQCD} is the lattice MiniMOM (MM) scheme \cite{MM1,MM2} rescaled to the conventional $\MSbar$ scale, i.e., $\Lambda_{MM} \mapsto {\overline {\Lambda}}$, i.e., in the leading order it coincides with the $\MSbar$ scheme, but has different scheme ($\beta$) coefficients $\beta_n$ ($n \geq 2$).}, because in that scheme the IR-safe (and holomorphic) QCD coupling was constructed $a(Q^2) \mapsto \A(Q^2)$ \cite{4l3dAQCD}, which at high $Q^2$ practically coincides with the underlying pQCD coupling $a(Q^2)$ (in LMM), reproduces the correct semihadronic $\tau$-decay  ratio $r_{\tau} \approx 0.20$, and behaves as $\A(Q^2) \sim Q^2$ when $Q^2 \to 0$ as suggested by large-volume lattice data on gluon and ghost propagator dressing functions in the Landau gauge \cite{LattcouplNf0,LattcouplNf24,Latt3gluon}. This QCD variant is called $3\delta$ $\A$QCD, because the spectral (discontinuity) function $\rho_{\A}(\sigma) \equiv {\rm Im} \A(Q^2= -\sigma - i \epsilon)$ in the low-$\sigma$ regime ($0 \leq \sigma \lesssim 1 \ {\rm GeV}^2$) is parametrized by three Dirac-delta functions, while $\rho_{\A}(\sigma)$  for higher $\sigma$ coincides with its underlying pQCD version $\rho_a(\sigma)$. The reason that the Borel transform (\ref{BtD4P}) is made in a renormalization scheme where a known holomorphic IR-safe QCD coupling $\A(Q^2)$ is available, will become clear in the next Section \ref{subsec:CharFunc}.

The parameter $\tal$, appearing at the $u=2$ ``pole term with $k=0$ multiplicity'' in Eq.~(\ref{BtD4P}), is not independent, because of the knowledge of the subleading part of the $D=4$ Wilson coefficient (we refer for details to \cite{GC1,BeJa08}). In the LMM scheme, the obtained value is $\tal_{\rm LMM} =  -0.14 \pm 0.12$.

After fixing the four parameters in the Borel transform ansatz (\ref{BtD4P}), the reexpansion (\ref{BtD1}) of the Borel transform then predicts the next coefficient $\td_4$ (and thus $d_4$) in the LMM scheme; this when transformed to the $\MSbar$ scheme, gives
\be
d_4(\MSbar)_{\rm pred.}=338.2
\label{d4MSbarpr}
\ee
For comparison, the Adler function is constructed also in another renormalization scheme, called Lambert scheme: it has a given value of the $c_2$ parameter\footnote{We recall that the scheme parameters are: $c_n \equiv \beta_n/\beta_0$, for $n \geq 2$. For convenience, the leading scheme parameter $\Lambda$ here (and in the LMM scheme) is such that the scaling is in the $\MSbar$ convention; i.e., the scheme is characterized only by the parameters $c_2, c_3, \ldots$.}, and $c_n=c_2^{n-1}/c_1^{n-2}$ for $n \geq 3$. The $c_2=-4.9$ Lambert scheme was used in the construction of the $2\delta$ $\A$QCD model \cite{2dAQCD} which has a holomorphic and IR-safe coupling. In this $c_2=-4.9$ Lambert scheme, we can now require that the first four coefficients are the exact ones (in that scheme), and that $d_4$ coefficient corresponds to that obtained in the LMM case; therefore, now five parameters can be fixed, and the ansatz in the Lambert scheme is
\bea
\lefteqn{
  {\rm B}[\btD](u)^{\rm (5 P)} =
}
\nonumber\\&&
\exp \left( \tK u \right) \pi {\Big \{}
\td_{2,1}^{\rm IR} \left[ \frac{1}{(2-u)} + \tal (-1) \ln \left( 1 - \frac{u}{2} \right) \right]
\nonumber\\ &&
+ \frac{ \td_{3,2}^{\rm IR} }{(3 - u)^2} + \frac{ \td_{3,1}^{\rm IR} }{(3 - u)} + \frac{ \td_{1,2}^{\rm UV} }{(1 + u)^2} {\Big \}}.
\label{BtD5P}
\eea
We are interested in the Adler function in this $c_2=-4.9$ Lambert scheme, because in this scheme an IR-safe (and holomorphic) QCD coupling $\A(Q^2)$ was constructed \cite{2dAQCD}, which at high $Q^2$ practically coincides with the underlying pQCD coupling $a(Q^2)$ and reproduces the correct $r_{\tau} \approx 0.20$; however, at $Q^2 \to 0$ the coupling is nonzero, $0 < \A(0) < \infty$, in contrast with the aforementioned $3 \delta$ $\A$QCD coupling\cite{4l3dAQCD}. This QCD variant is called $2\delta$ $\A$QCD, because its spectral function $\rho_{\A}(\sigma) \equiv {\rm Im} \A(Q^2=-\sigma - i \epsilon)$ in the low-$\sigma$ regime is parametrized by two Dirac-delta functions.

For comparison, the mentioned five-parameter Borel transform can also be applied in the $\MSbar$ scheme (five-loop, with ${\bar c}_n=0$ for $n \geq 5$), in the same way, and the parameters are fixed.

The results are given in Table. The $\tal$ parameters are:  $\tal_{\rm LMM} =  -0.14 \pm 0.12$; $\tal_{\rm Lamb.}  = -0.10 \pm 0.14$;
$\tal_{\rm \MSbar} =  -0.255 \pm 0.010$.
\begin{table*}[hbt]
\setlength{\tabcolsep}{1.5pc}
\label{tab4P5P}
\begin{tabular*}{\textwidth}{@{}l@{\extracolsep{\fill}}r|rrrrr}
  scheme & $\tK$ & $\td_{2,1}^{\rm IR}$ & $\td_{3,2}^{\rm IR}$ &  $\td_{3,1}^{\rm IR}$ & $\td_{1,2}^{\rm UV}$ 
\\
\hline
LMM  & -0.770405 & -1.83066 & 11.0498 & -    & 0.00588513 
\\
Lamb. &  0.2228 & 4.74582 & -1.04837 & -5.89714 & 0.0276003 
\\
$\MSbar$ & 0.5190 & 1.10826 & -0.481538 & -0.511642 & -0.0117704

\end{tabular*}
\end{table*}

\subsection{Characteristic function of the Adler function}
\label{subsec:CharFunc}
           
The characteristic (or: distribution) function $F_D(t)$ of a spacelike observable $\bD(Q^2)$ is
usually defined as such a function of $t>0$ that
\be
\bD_{\rm res.}(Q^2) = \int_0^{+\infty} \frac{dt}{t} F_D(t) a(t Q^2)
\label{Dres1}
\ee
represents the (leading-twist) resummation of $\bD(Q^2)$. Taylor expansion of the coupling $a(t Q^2)$ in $\ln(t Q^2)$ around $\ln Q^2$ then implies that the moments of $F_D(t)$ are precisely the coefficients $\td_n$ appearing in the auxiliary quantity $\btD(Q^2)$
\be
(- \beta_0)^n \int_0^{+\infty} \frac{dt}{t} F_D(t) \ln^n \left( \frac{t}{\kappa} \right) = \td_n(\kappa),
\label{FDSR}
\ee
where $n=0,1,\ldots$. Using these relations, with $\kappa=1$, and the expansion (\ref{BtD1}) in powers of $u$ for the Borel transform ${\rm B}[\btD](u)$, one obtains
\be
{\rm B}[\btD](u) = \int_0^{+\infty} \frac{dt}{t} F_D(t) t^{-u}.
\label{BtDMell}
\ee
Hence ${\rm B}[\btD](u)$ is the Mellin transform of $F_D(t)$. The inverse Mellin then gives the characteristic function $F_D(t)$ in terms of ${\rm B}[\btD](u)$ (cf.~\cite{Neubert} for application in the large-$\beta_0$ (one-loop) context)
\be
F_D(t) = \frac{1}{2 \pi i} \int_{1- i \infty}^{1 + i \infty} d u \; {\rm B}[\btD](u) t^u,
\label{FDinvMell}
\ee
For the Borel transforms (\ref{BtD4P}) and (\ref{BtD5P}), this inverse Mellin transform can be performed explicitly \cite{GC1}, and the result has the form
\bea
\lefteqn{
  \bD(Q^2)_{\rm res} =
}
\nonumber\\ &&
\int_0^1 \frac{dt}{t} G_D^{(-)}(t) a(t e^{-\tK} Q^2)
\nonumber\\ &&
+ \int_1^{\infty} \frac{dt}{t} G_D^{(+)}(t) a(t e^{-\tK} Q^2)
\nonumber\\ &&
+ \int_0^1 \frac{dt}{t} G_D^{\rm (SL)}(t) \left[ a(t e^{-\tK} Q^2) - a(e^{-\tK} Q^2) \right],
\label{Dres2b}
\eea
where the (characteristic) functions $G_D^{(\pm)}(t)$ and $G_D^{\rm (SL)}(t)$ involve the parameters of the mentioned Borel transforms (\ref{BtD4P}) and (\ref{BtD5P}), and powers of $t$ and $\ln t$, cf.~\cite{GC1}.

\subsection{Numerical evaluation}
\label{subsec:numev}

If the running coupling $a(Q^{'2})$ is holomorphic (analytic) in the complex $Q^{'2}$-plane excluding the timelike axis [$a(Q^{'2}) \mapsto \A(Q^{'2})$]\footnote{This means holomorphic in the generalized spacelike regime,  $Q^{'2} \in  \mathbb{C} \backslash (-\infty, -M_{\rm thr}^2]$, where $M_{\rm thr} \lesssim 0.1$ GeV is a threshold scale comparable with the lightest meson mass.}, it is IR-safe (finite when $Q^{'2} \to 0$), and thus the integration Eq.~(\ref{Dres2b}) can be performed. The problem of analyticity of QCD running couplings was addressed systematically already in the nineties \cite{ShS,MS,Sh1Sh2}, with a QCD variant called Analytic Perturbation Theory (APT) (for extensions and reviews, cf.~\cite{BMS,reviews}). Several versions of QCD holomorphic couplings have been applied in evaluations of various QCD quantities \cite{APTappl1,APTappl2,APTappl3,Luna}.\footnote{Yet another approach is to apply the requirement of the holomorphic behavior directly to QCD spacelike observables, cf.~Refs.~\cite{MagrGl,DeRafael,mes2,MagrTau,Nest3a,Nest3b}.}

  Two recently constructed QCD variants with holomorphic couplings $\A(Q^{'2})$, the aforementioned $2 \delta$ $\A$QCD \cite{2dAQCD} and $3 \delta$ $\A$QCD \cite{4l3dAQCD}, fulfill several phenomenological constraints of the low-$Q^{'2}$ QCD ($|Q^2| \lesssim 1 \ {\rm GeV}^2$) as mentioned earlier. The integrals in Eq.~(\ref{Dres2b}) can be performed in both variants ($a \mapsto \A$) without ambiguity because of the IR-safety of such couplings.

On the other hand, in pQCD in the usual schemes such as $\MSbar$, the running coupling $a(Q^{'2})$ is not holomorphic and not IR safe; it  has Landau singularities for positive small values of $Q^{'2}$, which makes the evaluation of the integrals in  (\ref{Dres2b}) ambiguous. To avoid this ambiguity, one may take the generalized principal value of these integrals, i.e., the integration is slightly shifted above the real positive axis, $a(t e^{-\tK} Q^2) \mapsto a(t e^{-\tK} Q^2+ i \epsilon)$, and the real part of the result is taken. Taking instead the imaginary part and dividing by $\pi$  [$\pm (1/\pi) {\rm Im} \ldots$] gives us a measure of ambiguity of such a result.

The results of this evaluation, for positive values of $Q^2$, are presented in Fig.~1.
 \begin{figure}[htb]
\centering\includegraphics[width=75mm,height=40mm]{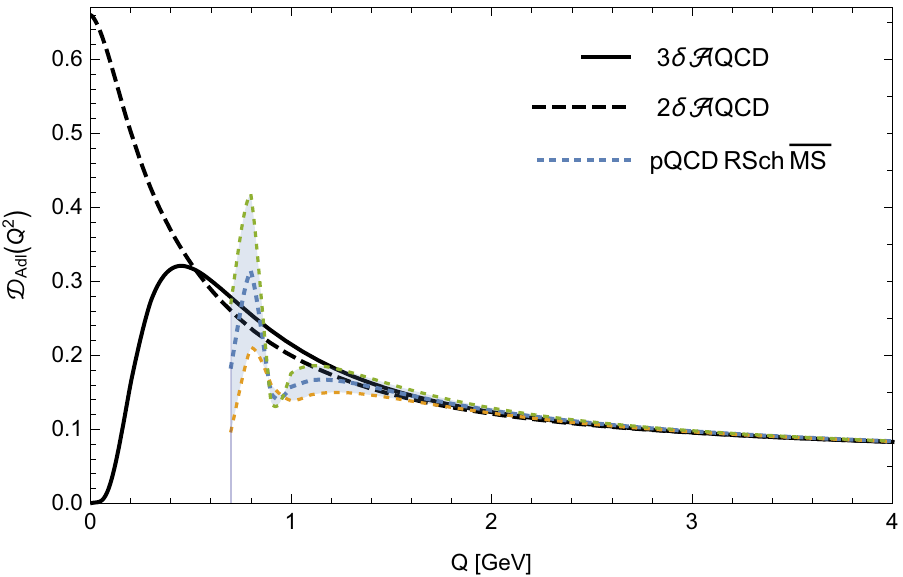}
\caption{The radiative Adler function resummed with the characteristic function according to Eq.~(\ref{Dres2b}) (where $a \mapsto \A$), as a function of $Q \equiv \sqrt{Q^2}$, for positive $Q^2$: in  $3\delta$ $\A$QCD (in the LMM renormalization scheme), and $2\delta$ $\A$QCD (in the Lambert $c_2=-4.9$ renormalization scheme). Included for comparison is the resummed pQCD Adler function $\bD(Q^2)_{\rm pQCD res}$ in the (five-loop) $\MSbar$ scheme, using modification of Eq.~(\ref{Dres2b}) as described in the text. All the three frameworks correspond to $\alpha_s(M_Z^2;\MSbar)=0.1185$.}
\label{FigDComb}
 \end{figure}
When the two holomorphic versions of QCD are applied, the results are regular down to $Q^2=0$, while the $\MSbar$ pQCD result is getting unstable and increasingly ambiguous at $Q < 1.5$ GeV. We also note that the two holomorphic results in the Figure start differing at $Q < 0.5$ GeV; this is so because the $2 \delta$ $\A$QCD coupling $\A(Q^2)$ tends to a positive finite value when $Q^2 \to 0$, and the  $3 \delta$ $\A$QCD coupling tends to zero (as $~\sim Q^2$) when $Q^2 \to 0$.

\subsection{$\tau$ decay ratio}

This method of evaluation of the Adler function (at general complex $Q^2$) can be used to evaluate the semihadronic (strangeless and massless) $\tau$ decay ratio $r_{\tau}^{(D=0)}$
\bea
\lefteqn{r_{\tau}^{(D=0)} = }
\nonumber\\ &&
\frac{1}{2 \pi} \int_{-\pi}^{\pi} d \theta (1 + e^{i \theta})^3 (1 - e^{i \theta}) \bD_{\rm Adl}( m_{\tau}^2 e^{i \theta} ),
\label{rtau1}
\eea
where the subscript $D=0$ denotes the leading-twist (dimension zero) contribution. In 3$\delta$ $\A$QCD, in which the truncated sum in logarithmic derivatives,
$\bD(Q^2) = \sum_0^3 \td_n \tA_{n+1}(Q^2)$, gave $r_{\tau}^{(D=0)}=0.201$, we obtain with the resummation method: $r_{\tau}^{(D=0)}=0.2056$.
In 2$\delta$ $\A$QCD, in$r_{\tau}^{(D=0)}$ which LB+bLB approach (bLB truncated at $\tA_4$) gave $r_{\tau}^{(D=0)}=0.201$, we obtain with the resummation method: $r_{\tau}^{(D=0)}=0.1973$. We see that the resummation does contribute nonnegligible terms, in comparison with the usual truncated approaches. 

\section{Summary}
\label{sec:Summ}

 \begin{itemize}
   
 \item  
A method of evaluation of spacelike QCD observables $\bD(Q^2)$ was developed, motivated by the renormalon structure of these quantities.

\item
A related auxiliary quantity $\btD(Q^2)$ was introduced, which is renomalization scale independent only at the one-loop level, and agrees with  $\bD(Q^2)$ at one-loop level.

\item
A large-$\beta_0$-type renormalon-motivated ansatz is made for the Borel transform ${\rm B}[\btD](u)$ of  $\btD(Q^2)$. This leads to a correctly ``dressed'' Borel transform ${\rm B}[\bD](u)$ of the considered observable $\bD(Q^2)$.

\item
Subsequently, a Neubert-type characteristic (distribution) function, $G_D^{(\pm)}(t)$ and  $G_D^{\rm (SL)}(t)$, is obtained for the considered observable $\bD(Q^2)$ as the inverse Mellin transform of the Borel transform of $\btD(Q^2)$.

\item
As an illustration, the method is applied to the massless Adler function and the related decay ratio of the $\tau$ lepton semihadronic decays.
\end{itemize}

\end{document}